      [1999/12/01 v1.4c Il Nuovo Cimento]
\documentclass{cimento}
\usepackage{graphicx}

\title{The MOLLER Experiment}
\author{J.~Mammei\from{ins:x} for the MOLLER Collaboration}
\instlist{\inst{ins:x} University of Massachusetts, Amherst}

\PACSes{
\PACSit{12.15.-y}{Electroweak Interactions}
\PACSit{12.15.-Mm}{Neutral Currents}
\PACSit{11.30.Er}{Parity Violation}
}
\begin{document}

\maketitle

\begin{abstract}
The MOLLER experiment will measure the weak charge of the electron, $Q^e_W = 1 - 4\sin^2\theta_W$, with a precision of 2.3\% by measuring the parity-violating asymmetry in electron-electron (M\o ller) scattering.  This measurement will provide an ultra-precise measurement of the weak mixing angle, $\sin^2\theta_W$, which is on par with the two most precise collider measurements at the Z$^0$-pole.  The precision of the experiment, with a fractional accuracy in the determination of $\sin^2\theta_W$ $\approx 0.1$\%, makes it a probe of physics beyond the Standard Model with sensitivities to mass scales of new physics up to 7.5~TeV.

\end{abstract}

\section{Introduction}

\begin{figure}[htp]
\begin{center}
\begin{tabular}{cccc}
\includegraphics[width=3.1cm]{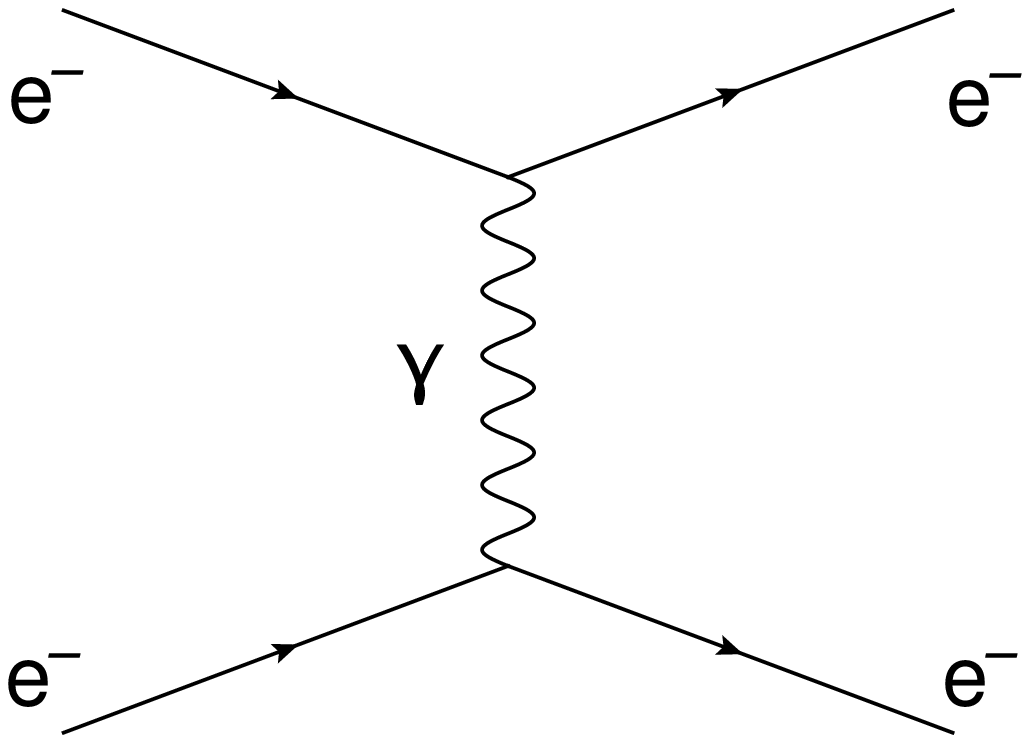}&
\includegraphics[width=3.1cm]{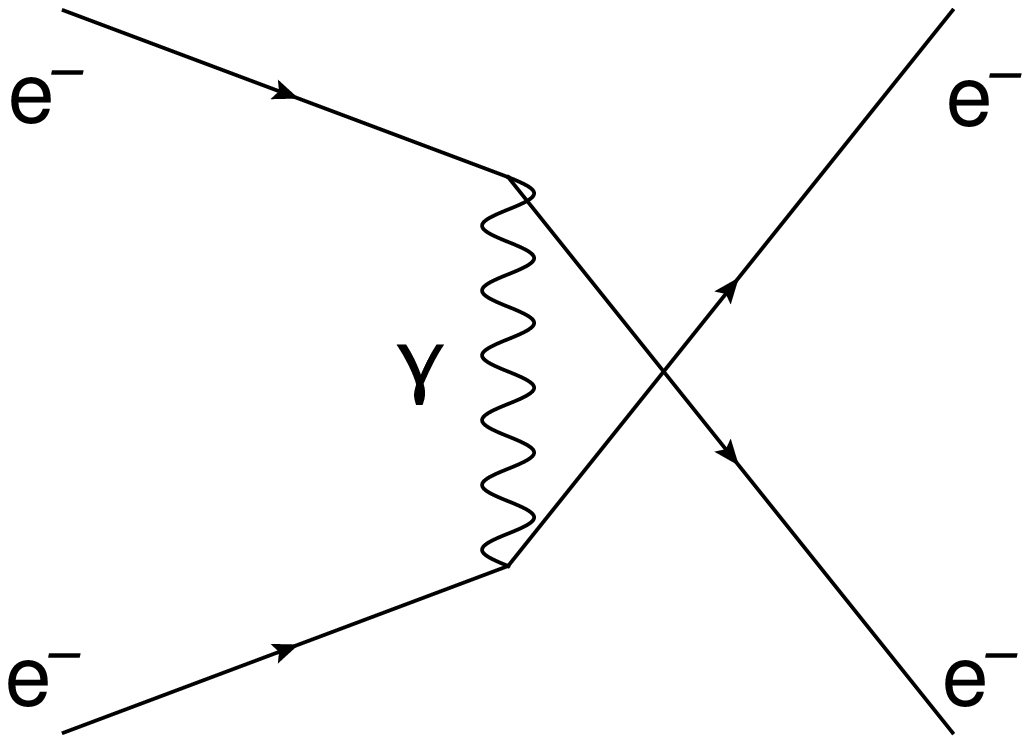}&
\includegraphics[width=3.1cm]{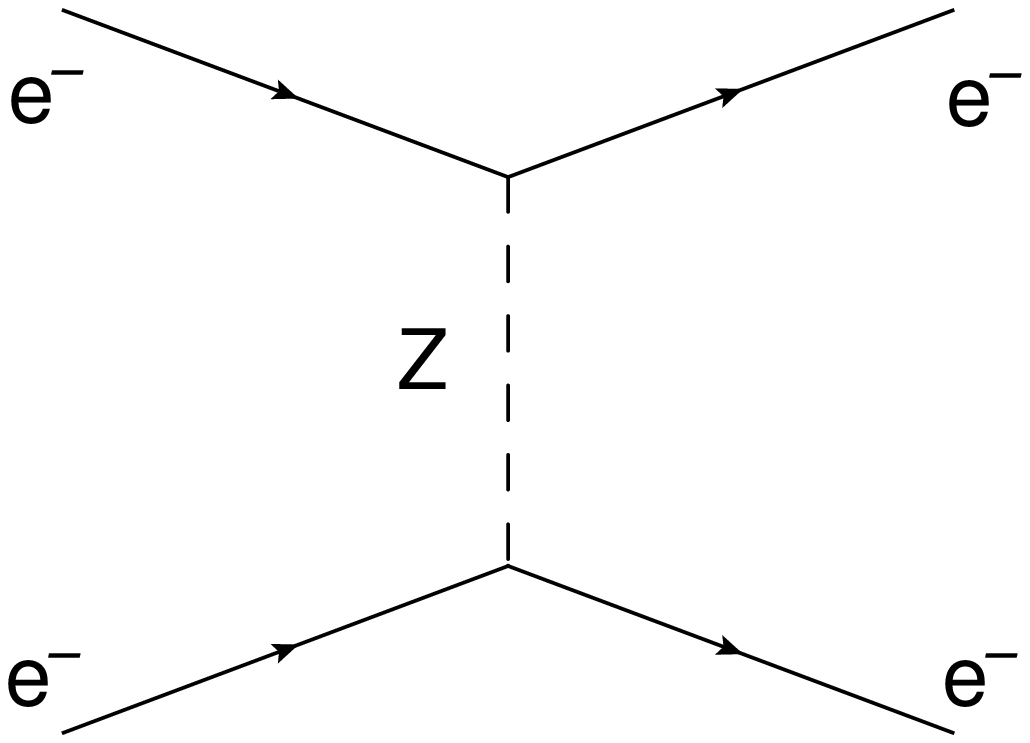}&
\includegraphics[width=3.1cm]{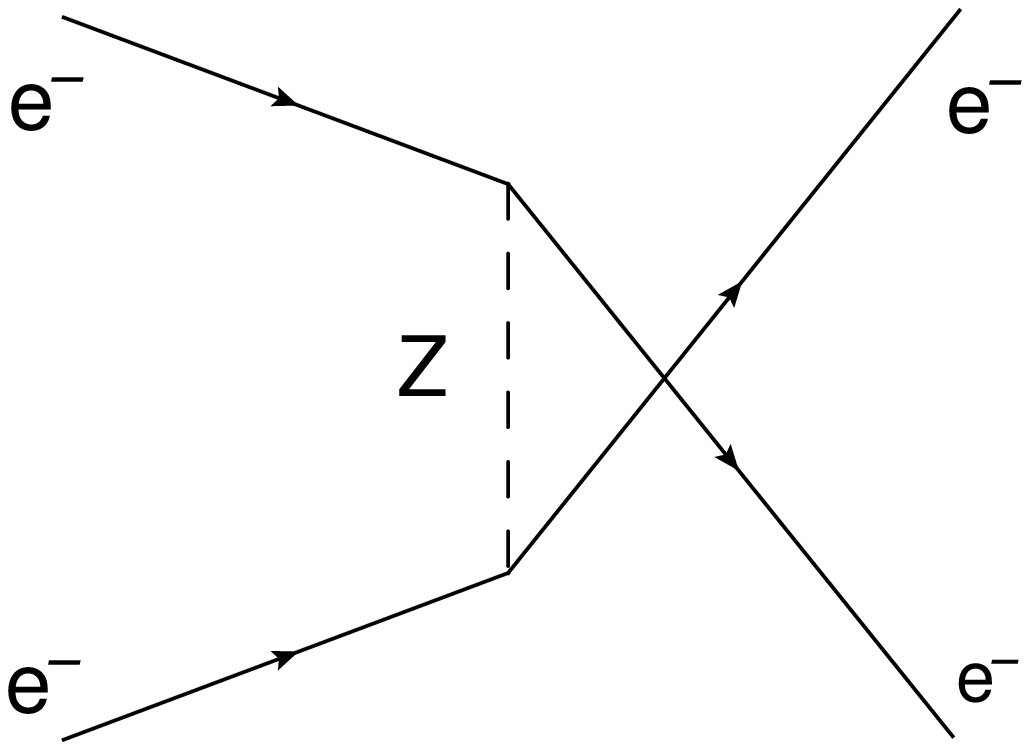}
\end{tabular}
\end{center}
\caption{{\em Feynman diagrams for M\o ller scattering at tree level (reproduced from Ref.~\cite{Czarnecki:2000ic})}}
\label{figtree}
\end{figure}

The Standard Model (SM) summarizes our knowledge of fundamental particles and their interactions.  Polarized electron scattering off unpolarized targets provides a clean window to study weak neutral current interactions by measuring the parity-violating asymmetry, $A_{PV}$.  The MOLLER experiment will test the SM prediction of the weak mixing angle, $\sin^2\theta_W$, by measuring $A_{PV}$ in polarized electron scattering from atomic electrons (M\o ller scattering).  The $A_{PV}$ is defined by
\begin{equation}
A_{PV} = {\sigma_R-\sigma_L\over\sigma_R+\sigma_L} \; , \label {eq:adef}
\end{equation}
where $\sigma_R$ ($\sigma_L$) is the scattering cross section of incident right- (left-) handed electrons.  At four-momentum transfers much smaller than the mass of the Z$^0$ boson, $Q^2\ll M_Z^2$, $A_{PV}$ is dominated by the interference between photon and $Z^0$ boson exchange~\cite{zeld}.  The leading order Feynman diagrams relevant for M\o ller scattering, which involve both direct and exchange diagrams that interfere with each other, are shown in fig.~\ref{figtree}. The resulting asymmetry is given by~\cite{Derman:1979zc}  
\begin{equation}
A_{PV} = mE{G_F\over\sqrt{2}\pi\alpha}{4\sin^2\theta\over(3+\cos^2\theta)^2}Q^e_W = mE{G_F\over\sqrt{2}\pi\alpha}\frac{2y(1-y)}{1+y^4+(1-y)^4}Q^e_W \; ,
\label {eq:amoll}
\end{equation}
where $\alpha$ is the fine structure constant, $G_F$ the Fermi constant, $E$ the incident beam energy, $m$ the electron mass, $\theta$ the scattering angle in the center of mass frame, and $y\equiv 1-E^\prime/E$ where $E^\prime$ is the energy of one of the scattered electrons.  

Within the SM the weak charge of the electron, $Q^e_W$, is proportional to the product of the electron's vector and axial-vector couplings to the $Z^0$ boson, and the weak neutral current amplitudes are functions of the weak mixing angle $\sin^2\theta_W$.  The world average of the two most precise independent determinations of $\sin^2\theta_W$ is consistent with other electroweak measurements and constraints on the Higgs boson mass $M_H$, but the measurements actually differ by over 3 standard deviations.  Choosing one or the other central value ruins this consistency and implies very different new high-energy dynamics.  We propose to measure $\sin^2\theta_W$ to a sensitivity of $\delta(\sin^2\theta_W) = \pm 0.00029$, the only method available in the next decade to directly address this issue at the same level of precision and interpretability.  

MOLLER will also be {\it the} most sensitive probe of new flavor and CP-conserving neutral current interactions in the leptonic sector, sensitive to interaction amplitudes as small as $1.5\times 10^{-3}$ times the Fermi constant.  New neutral current interactions are best parameterized model-independently at low energies by effective four-fermion interactions.  Focusing on vector and axial-vector interactions between electrons and/or positrons, such an interaction Lagrangian takes the form~\cite{Eichten:1983hw}:
\begin{equation}
{\cal L}_\mathrm{e_1e_2} = \sum_{i,j = L,R} {g_{ij}^2\over 2\Lambda^2} {\bar e_i}\gamma_\mu e_i{\bar e_j}\gamma^\mu e_j \; ,
\label{contact}
\end{equation}
where $e_{L/R} = {1\over 2}(1 \mp \gamma_5)\psi_e$ are the usual chirality projections of the electron spinor, $\Lambda$ the mass scale of the new contact interaction and $g_{ij} = g_{ij}^*$ coupling constants, with $g_{RL} = g_{LR}$.  For the proposed measurement with 2.3\% total uncertainty (and no additional theoretical uncertainty) the resulting sensitivity to new four-electron contact interaction amplitudes $g^2_{RR}-g^2_{LL}$ is $\sim$7.5~TeV.  The proposed measurement will greatly extend the current sensitivity of four-electron contact interactions, both qualitatively and quantitatively, and is complementary to direct searches. 

\begin{figure}[ht]
\begin{minipage}[b]{0.49\linewidth}
\centering
    \includegraphics[width=2.5in]{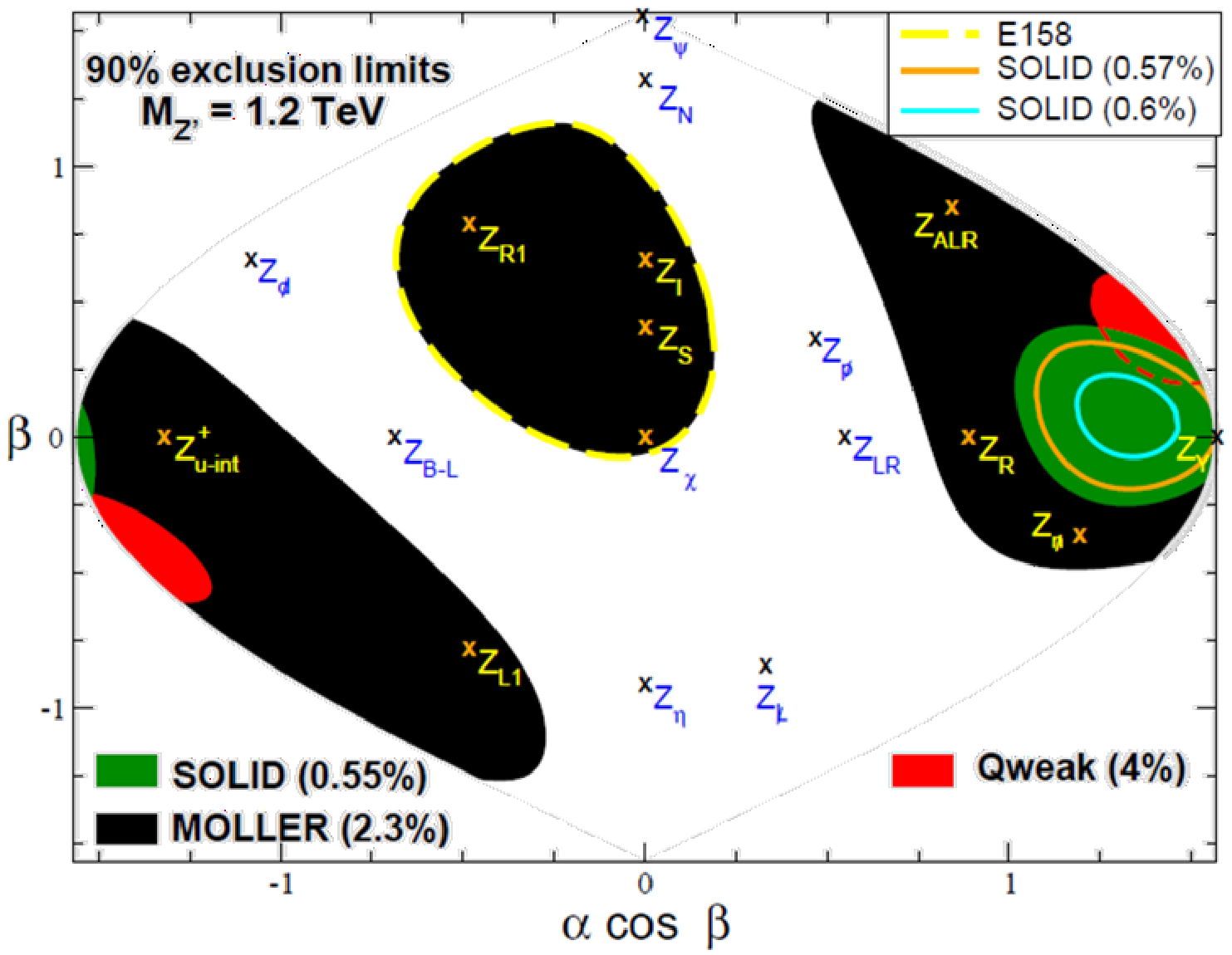}
      \caption{90\% C.L.\ exclusion regions for a 1.2~TeV $Z^\prime$ from the $E_6$ gauge group for E158, and assuming future experiments measure the SM value.}
  \label{zprimemollerreach}
\end{minipage}
\hspace{0.3cm}
\begin{minipage}[b]{0.49\linewidth}
\centering
\includegraphics[width=2.5in]{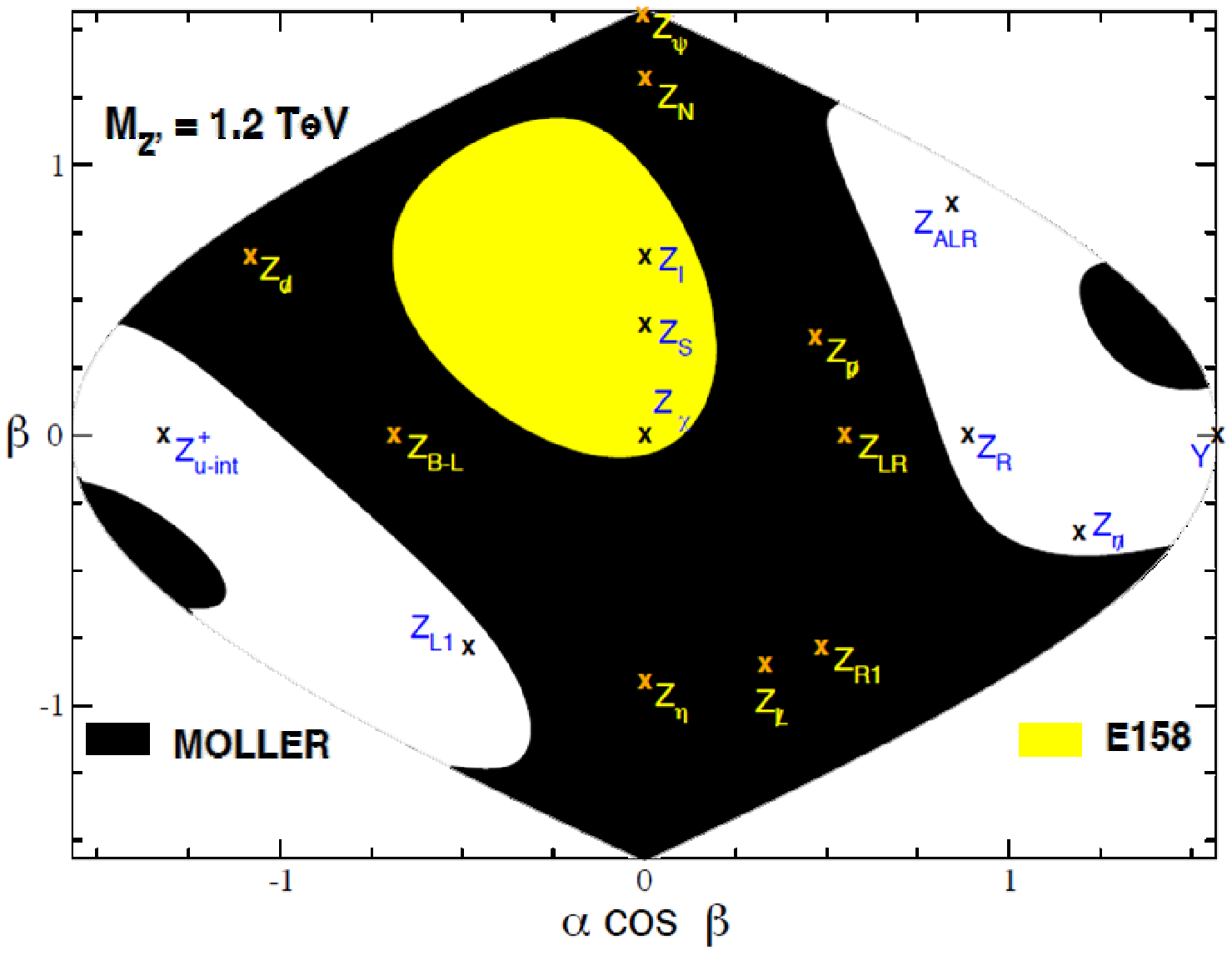}
\caption{90\%\ C.L. exclusion regions for a 1.2~TeV $Z^\prime$ from the $E_6$ gauge group for E158, and assuming the MOLLER value is half-way between E158 and the SM.}
\label{zprimemollere158}
\end{minipage}
\end{figure}

It is straightforward to examine the reach MOLLER will have in specific models (the mass scale depends on the size of the coupling)~\cite{RamseyMusolf:1999qk}.  As a specific example, a comprehensive analysis of the MOLLER sensitivity to TeV-scale $Z^\prime$ bosons has recently been carried out~\cite{newjenszprime} for a large class of models contained in the $E_6$ gauge group, where $Z^\prime$ bosons with the same electroweak charges as SM particles are motivated because they arise in many superstring models as well as from a bottom-up approach~\cite{Erler:2000wu}.  These models are spanned by two parameters $\alpha$ and $\beta$ in the range $\pm\pi/2$.  $\alpha = 0$ corresponds to the $E_6$ models considered for example in Ref.~\cite{ngpaper}, while $\alpha \neq 0$ can be interpreted as non-vanishing kinetic mixing, assuming that this kinetic mixing has been undone by field re-definitions.  $\beta = 0$ correspond to $SO(10)$ models, which include models based on left-right symmetry.

The MOLLER reach for a 1.2 TeV $Z^\prime$ from this model class, assuming the value predicted by the SM is measured, is shown in fig.~\ref{zprimemollerreach}, while fig.~\ref{zprimemollere158} shows the allowable region assuming MOLLER instead measures a value half-way between the SM value and the E158 central value.  The current region excluded by E158 is shown in both figures in yellow.  Thus, the combination of potential MOLLER and LHC anomalies would point to a small list of $Z^\prime$ models, whose effects on other precision electroweak observables can then be further explored at LHC and elsewhere.  

However, if MOLLER measures a central value consistent with E158, then a clear violation of the SM would be established at more than 5$\sigma$.  We note that no $Z^\prime$ from the above mentioned model class could explain such a MOLLER deviation.

\section{Experimental Overview}

The MOLLER experiment will run in Hall A at Jefferson Laboratory, making use of the 11 GeV longitudinally polarized ($\sim$85\%) electron beam, generated via photoemission on a GaAs photocathode by circularly polarized laser light.  The target is a 1.5~m liquid hydrogen target capable of dissipating 5~kW of beam power.  M\o ller electrons in the full range of the azimuth (achieved by using an odd number of coils and identical particle scattering) and polar angles 5 mrad $<\theta_{lab}<$ 17 mrad, will be separated from background and focused $\sim 30$~m downstream of the target by a spectrometer system consisting of a pair of toroidal magnet assemblies and precision collimators.  The upstream magnet is a traditional resistive toroidal magnet, while the downstream magnet has a novel shape designed to focus the large range of scattered electron angles and energies.  
\begin{figure}[ht]
\centering
\includegraphics[width=3in]{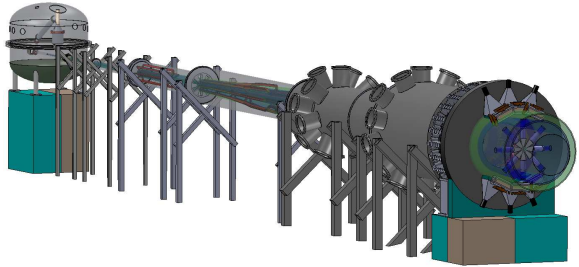}
\caption{CAD model of the conceptual layout of the experiment, looking upstream.  The target chamber is located in the upper left, then the upstream and hybrid toroids, Roman pots for the tracking detectors, and the main detector stand, with background detectors at the far downstream end in the lower right.}
\label{figCADMOLLER}
\end{figure}

The M\o ller electrons will be incident on a system of quartz detectors in which the resulting Cerenkov light will provide a relative measure of the scattered flux.    Rapid polarization (helicity) reversal of the electrons will be used to suppress spurious systematic effects.  $A_{PV}$ will be extracted from the fractional difference in the integrated Cerenkov light response between helicity reversals.  Additional systematic suppression below the part-per-billion (ppb) level will be accomplished by periodically reversing the sign of the physics asymmetry.  We plan to introduce this ``slow helicity reversal" with three independent methods: the introduction of an additional half-cycle $g-2$ rotation to the electrons in the recirculating arcs of the accelerator, the use of an insertable half-wave plate in the injector, and a full flip of the beam polarization direction with the aid of two Wien rotators and a solenoid lens (the ``Double-Wien''). 

 \begin{table}
 \caption{Summary of projected fractional systematic errors on the measurement of $Q^e_W$.  The fractional statistical error is 2.1\%.}\label{tab:molsyst}
 \begin{center}
 \begin{narrowtabular}{2cm}{lc}
    Error Source & Fractional Error (\%) \\
    \hline
    absolute value of $Q^2$ & 0.5 \\
    beam (second order) & 0.4 \\
    beam polarization & 0.4 \\
    $e + p (+\gamma)\rightarrow e+X(+\gamma)$ & 0.4 \\
    beam (position, angle, energy) & 0.4 \\
    beam (intensity) & 0.3 \\
    $e + p (+\gamma)\rightarrow e+p(+\gamma)$ & 0.3 \\
    $\gamma^{(*)} + p \rightarrow \pi + X$ & 0.3 \\
    transverse polarization & 0.2 \\
    neutrals (soft photons, neutrons) & 0.1 \\
    \hline
    Total  systematic              &  {\bf 1.1} \\
\end{narrowtabular}
\end{center}
\end{table}

Simultaneously with data collection, fluctuations in the electron beam energy and trajectory and their potential systematic effects on $A_{PV}$ will be precisely monitored, active feedback loops will minimize beam helicity correlations, and detector response to beam fluctuations will be continuously calibrated.  Background fractions and their helicity-correlated asymmetries will be measured by dedicated auxiliary detectors. The absolute value of $Q^2$ will be calibrated using tracking detectors. The electron beam polarization will be measured continuously by two independent polarimeter systems.  The predicted value of $A_{PV}$ for the proposed experimental design is $\sim 35$~ppb and our goal is to measure this quantity with a precision of 0.73 ppb.  We tabulate our estimates of the most important systematic errors in Table~\ref{tab:molsyst}.  

\section{Experiment Status and Plans}

We are proposing to take data in three separate run periods to ensure that important technical milestones are met and that each run will provide publishable results that will significantly add to our knowledge of electroweak physics to date.  While the MOLLER apparatus is being designed for a beam current of 85~$\mu$A at 11 GeV, we have assumed a beam current of 75 $\mu$A and a beam polarization of 80\%\ in formulating the run plan.  If higher beam current and/or higher beam polarization are considered routine, the amount of time needed to run could correspondingly be reduced using the appropriate $P^2I$ factor.   

In order to estimate the time needed to reach the desired statistical accuracy it is necessary to take into account both the overall efficiency and how close one can approach counting statistics in an instantaneous raw asymmetry measurement.  The overall efficiency depends on the efficiency of the apparatus itself and the accelerator efficiency, which we estimate as 90\% and 70\% respectively, for an overall efficiency of 60\%.  At 960 Hz, the width of the measured asymmetry per pulse pair, $\sigma(A_i)$, is 83 ppm, but this width depends on the sources of additional fluctuations.  However, we are unlikely to achieve the best efficiency or asymmetry width at first, so we assumed total efficiencies of 40, 50 and 60\%\, respectively, and asymmetry widths of 100, 95 and 90 ppm, respectively, for the three running periods.  This leads to run periods of 11, 30 and 60 weeks (including commissioning) with expected statistical errors of 11\%, 4.0\% and 2.4\%, respectively.

MOLLER is a fourth generation parity-violation experiment at Jefferson Laboratory.  Apart from the obvious challenge of measuring a raw asymmetry with a statistical error less than 1 ppb, an equally challenging task is to calibrate and monitor the absolute normalization of $A_{PV}$ at the sub-1\%\ level.  The collaboration continues to gain extensive experience on all aspects of such measurements as work continues on executing the third generation experiments PREX~\cite{PREX} and Qweak~\cite{Qweak}.  Simulation and design of the various aspects of MOLLER are ongoing, and is expected to take two to three years.  Work is also being done to improve beam transport and instrumentation.  In addition, upgrades to the Compton and M\o ller polarimeters are being planned.  It is envisioned that construction and assembly will take three years, to be followed by three data collection periods with progressively improved statistical errors and systematic control over a subsequent three to four year period.  
 

\acknowledgments
Thanks to the organizers for a wonderful meeting.  This work is funded in part by US Department of Energy Grant No. DE-FG02-88R40415-A018.

\end{document}